\documentclass[twocolumn,showpacs,preprintnumbers,amsmath,amssymb]{revtex4}

\usepackage{graphicx}
\usepackage{dcolumn}
\usepackage{bm}
\usepackage{subfigure}
\usepackage{placeins}
\usepackage{multirow}
\usepackage{amsmath}
\usepackage{slashbox}
\usepackage{multirow}
 
\begin{document}
 

\title{Stability of Climate Networks with Time}

\author{Y. Berezin}
\affiliation{Minerva Center and Department of Physics, Bar Ilan University, Ramat Gan, Israel.}

\email{bereziny@google.com}
\author{A. Gozolchiani}%
\affiliation{Minerva Center and Department of Physics, Bar Ilan University, Ramat Gan, Israel.}
\author{S. Havlin}%
\affiliation{Minerva Center and Department of Physics, Bar Ilan University, Ramat Gan, Israel.}



\date{\today}

\begin{abstract}
   We construct and analyze climate networks based on daily satellite
   measurements of temperatures and geopotential heights. We show that these
   networks are stable during time and are similar over different altitudes.
   Each link in our network is stable with typical $15\%$ variability. The
   entire hierarchy of links is about $80\%$ consistent during time. We show that about
   half of this stability is due to the spatial 2D embedding of the network, and
   half is due to physical coupling mechanisms. The network stability of
   equatorial regions is found to be lower compared to the stability of a
   typical network in non--equatorial regions.
\end{abstract}


\maketitle
 
\section{\label{sec:level1}Introduction}

 	During the past decade, methods from network theory have been applied to
	describe complex systems that are composed of many
	interacting components (see e.g.
	~\cite{Barabási15101999, barabasi, Newman2006, Dorogovtsev2003,
	Newman03thestructure, watts_strogatz, Stewart, airports, bootstrap,
	bararev, pastor2004evolution, caldarelli07, caldarelli2007large,
	barrat2008dynamical, newman2010networks, cohen2010complex, Albert2000,
	Barthelemy2011, callaway, acebron, hiroshi, fireflies, tamara,cohen_1,
	song, neurochip}). While in some cases the representation
	of such systems as a network is obvious and the nodes and links are identified
	directly (e.g. cables connecting computers in a computer
	network)~\cite{barabasi}, in many real world networks such as biological
	systems~\cite{Dorogovtsev2003}, neural networks~\cite{neurochip}, climate
	networks~\cite{survival,EPLgozyamga,EBauton,tsoniselnino,donges_mi,donges,Donges2011,Guez2011,Tsonis2011}
	and	others, the identification of links is not direct and it is based on
	statistical analysis of the similarity of the dynamics of
	nodes~\cite{Pikovsky_Rosenblum_Kurths_2003}.
   
    In recent years it was suggested that climate fields such as
    temperature and geopotential height at a certain pressure
    level can be represented as a climate network where the nodes
    are geographical sites and the links are the information flow between these
    sites (nodes).
	Although the dynamics within a single node is unpredictable and chaotic, yet
	the dynamic of one node may be coupled to the dynamics of other nodes and
	could be observed. The correlations between the dynamics in two different nodes is
	represented in our network as a link between them. Since we deal with
	climatological data, the correlations (Sec.~\ref{sec:construction_method})
	might be with a time-delay. Therefore, each link is quantified by two
	parameters; a strength value which quantifies the intensity of the cross
	correlations, and a time delay value which quantifies the delay in the data flow
	between the two nodes.
\section{\label{sec:level}The climate network}

\subsection{\label{sec:level2}Data}
\begin{figure}[ht]
	\includegraphics[width = 0.32\textwidth,angle=270]{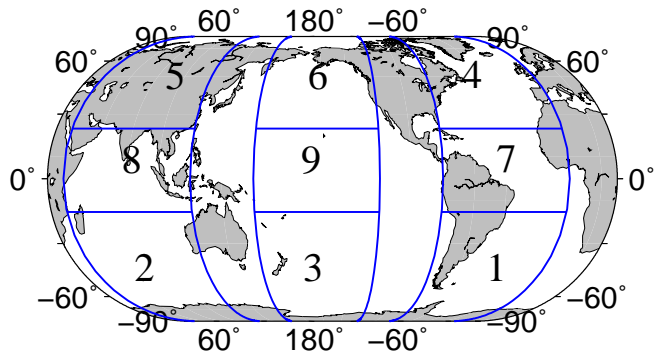}
    \caption{(color online). The geographical locations of the 9 separate zones,
    on which we base our network analysis. \label{Zones}}
\end{figure}

    We analyze data obtained from a reanalysis project~\cite{reanalysis2}.
    The records consist of the reanalysis air temperature field and
    the geopotential height field, for the 1000hPa, 925hPa, 850hPa, 700hPa,
    500hPa and 300hPa isobars. We use daily values between the years 1948-2006.
    The data is arranged on a world-wide grid with a resolution of $5^\circ\times5^\circ$.
   	We divide the globe into 9 zones (see Fig.~\ref{Zones}), in order to
   	identify different network dynamics specific to different zones.

\subsection{\label{sec:construction_method}The network construction method}
\begin{figure}[ht]
		\includegraphics[width = 8cm,height = 5cm]{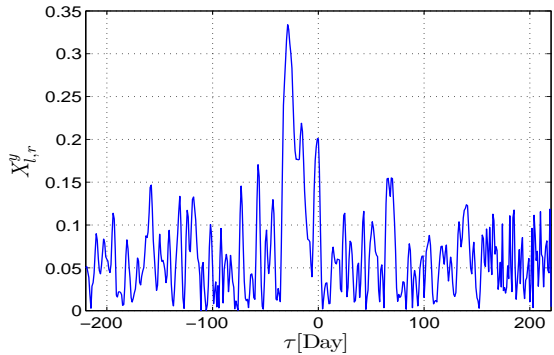}
     	\caption{A typical cross-covariance function between two sites,
     	representing the level of correlation within a time lag ranging from
     	$-\tau_{max}$ to $+\tau_{max}$ where $\tau_{max} = 220$. In the current
     	example $W^{y}_{l,r} = \frac{0.3-0.05}{0.05}=5$, $T^{y}_{l,r} = -15$.
     	\label{corr_func}}
	\end{figure}

	We analyze daily climatological records (temperature/pressure) taken from a
	grid in various geographical zones (Fig.~\ref{Zones}). To avoid the trivial
	effect of seasonal trends we subtract from the records of each day the yearly
	average value of that day. Specifically, we take the climatological
	signal (temperature/pressure) of a given site in the grid to be
	$\widetilde{S}^y(d)$, where $y$ is the year and $d$ is the day (ranging from
	$1$ to $365$) of that year. The new signal will be
	$S^y(d)=\widetilde{S}^y(d)-\frac{1}{N}\sum_y\widetilde{S}^y(d)$, where $N$ is
	the number of years available in the record. For each pair of sites $l$ and $r$ in
	a specific zone, we compute the absolute value of the cross-covariance function
	$X^{y}_{l,r}$ of their local climatological signals such as
	temperature/pressure in the range of time delays
	$\tau\in\left[-\tau_{max},\tau_{max}\right]$ integrated over a specific year
	($y$) (see Fig.~\ref{corr_func}). Using the cross-covariance function
	$X^{y}_{l,r}$ we define the strength of the link to be
	$W^{y}_{l,r}=(MAX(X^{y}_{l,r})-<X^{y}_{l,r}>)/STD(X^{y}_{l,r})$, where
	$<\ldots>$, $MAX$ and $STD$ are the mean value, maximal value and the standard
	deviation of $X^{y}_{l,r}$ in the range of $\tau$, respectively. The matrix $W_{l,r}^y$ represents the
	{\it weighted adjacency matrix} of the network at year $y$. The time
	shift at which $X^{y}_{l,r}$ is maximal is defined as the link time delay,
	and denoted as $T^{y}_{l,r}$.

\section{\label{sec:Results}Results}
\subsection{\label{sec:single}Stability of single links}	

\begin{figure}[ht]
        \begin{center}
            \subfigure[\ Short distance link ($750Km$)]{\includegraphics[width =
            8cm,height = 5cm]{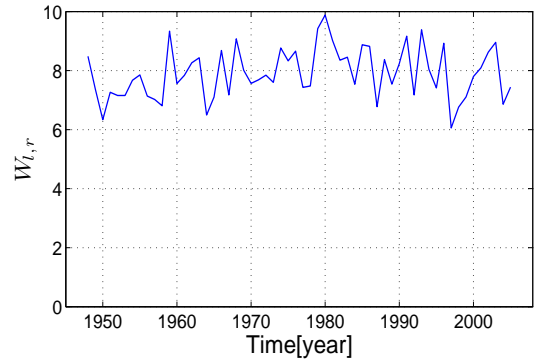}}
            \subfigure[\ Long distance link ($1500Km$)]{\includegraphics[width =
            8cm,height = 5cm]{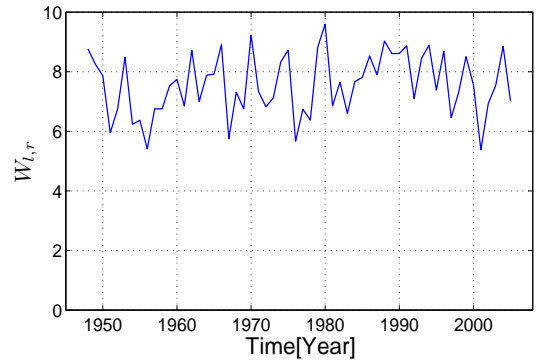}}
        \end{center}

        \caption{Two examples of typical dynamics of a link strength
        $(W)$ during the years. In (a) the distance between the two sites is
        about $750 Km$, the average time delay $\overline{T} = 0$ $(day)$ and
        the variation in the link strength, $STD(W_{l,r})/\overline{W_{l,r}}
        = 0.1$. In (b) the distance between the two sites is about $1500 Km$,
        the average time delay $\overline{T} = 1$ $(day)$ and the variation in the
        link strength, $STD(W_{l,r}) = 0.1$.
        \label{single_link}}
    \end{figure}

\begin{figure}[ht]
			
			\includegraphics[width=0.5\textwidth,height=5cm]{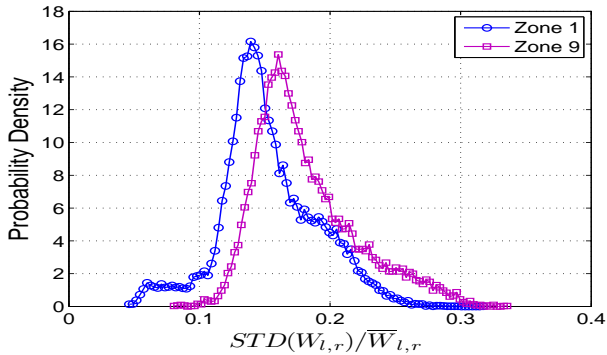}
			\caption{Distribution of the links variation during time,
			$STD\left(W\right)/\overline{W}$ in two zones. Zone $1$ and Zone $9$, both
			for network based on temperature at 850hPa isobar. 
    		\label{stdHist}}
    \end{figure}

    First we focus on the behavior of single links. We analyze the yearly
    variations of the link strength $W^y_{l,r}$. We find that a typical link
    maintains its strength $W_{l,r}$ during the years with typically small
   	fluctuations of $15\%$. This stability in the strength of the links behavior
   	is valid for links across long and short distances (see
   	Fig.~\ref{single_link}).
	In Fig.~\ref{stdHist} we show the distribution of the relative standard
	deviation of all links in two networks (a) the network located at Zone 1
	(a non--equatorial region) and (b) the network located at Zone 9 (an equatorial
	region) (see Fig.~\ref{Zones}). We observe a significant difference in this
	distribution between networks located in equatorial regions and networks located
	in non-equatorial regions. While in networks located in equatorial regions (zones
	7-9), the minimum variation is about $0.1$ and the maximum is about $0.3$, in   
	networks located in non--equatorial regions (zones
	1-6), the minimum variation is about $0.05$ and the maximum is about $0.25$
	(see Fig.~\ref{stdHist}). Therefore the link strengths, $W_{l,r}$,  in
	non--equatorial regions tend to be more stable than the link strengths,
	$W_{l,r}$,  in equatorial regions.


	A typical auto-correlation function for climatological records of a
	specific node decays rapidly as a power low with
	time~\cite{autocorrelation_decrease,
	autocorrelation_decrease_2,autocorrelation_decrease_5,
	autocorrelation_decrease_6, autocorrelation_decrease_3,
	Santhanam2005713}. Therefore, this stability of the links over many years is
	surprising and may suggest that we can extract new information from the links
	between the nodes.

\begin{figure}[ht]
		\subfigure[\ Zone 1]{\includegraphics[width = 0.5\textwidth,height =
    	5cm]{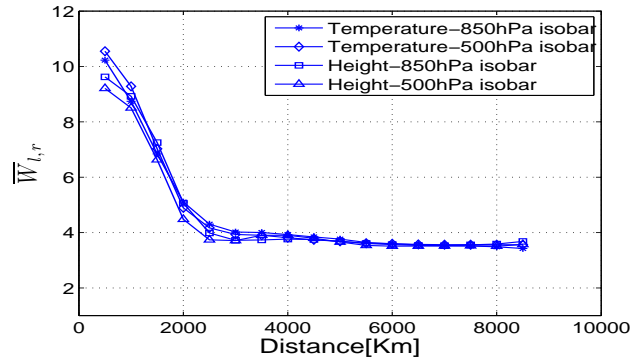}\label{distance_profile_eq_a}}
    	\subfigure[\ Zone 9]{\includegraphics[width =
    	0.5\textwidth,height = 5cm]{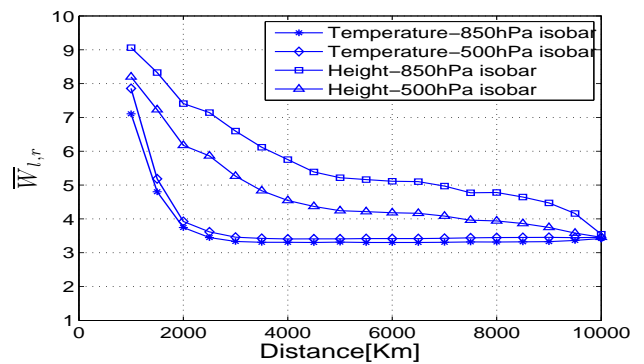}\label{distance_profile_eq_b}}
    	\caption{The dependence of $\overline{W}_{l,r}$ on $D_{l,r}$ in
    	two typical locations, (a) Zone 1 (non--equatorial region). (b) Zone 9
    	(equatorial region). The four curves describe four networks which are based
    	on geopotential height measurement at 850 and 500hPa isobar and based on
    	temperature measurement at 850 and 500hPa isobar.
    	\label{distance_profile_eq}}
    \end{figure}

	Analyzing the influence of spatial distances $D$, between the nodes on
	the strength $W_{D}$ of the link leads to the observation of a strong
	dependence of $W_{D}$ on $D$ (Fig.~\ref{distance_profile_eq}). Here
	$W_D \equiv \overline{W}^y_{l,r}$ is the average over all link strengths $W$ at
	distance $D$, and over all years, $y$. It is seen that for $D>2000~Km$, $W_D$
	reaches a low and almost constant value. This constant value can be regarded
	(as will be seen in Sec~\ref{sec:false_true}) as the level of noise. However,
	we observe a significant difference in this dependence between networks located in
	equatorial regions and networks located in non-equatorial regions (compare
	Fig.~~\ref{distance_profile_eq_a} and Fig.~\ref{distance_profile_eq_b}). In
	networks located in equatorial regions (zones 7-9), $W_{D}$ decreases
	significantly slower with time compared to other regions. This difference
	mainly appears in networks based on the geopotential height field.
	
\subsection{\label{sec:Stability}Stability of the entire network}

\begin{figure}[ht]
        \begin{center}
            \subfigure[\ Temp-850hPa, zone1]{\includegraphics[width = 8cm,height
            =
            5cm]{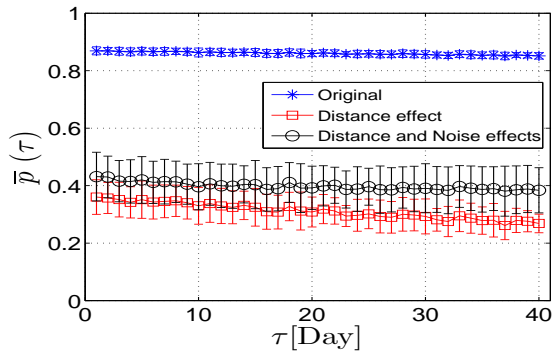}\label{tau_all_fillter_a}}

            \subfigure[\ Temp-850hPa, zone9]{\includegraphics[width = 8cm,height
            =
            5cm]{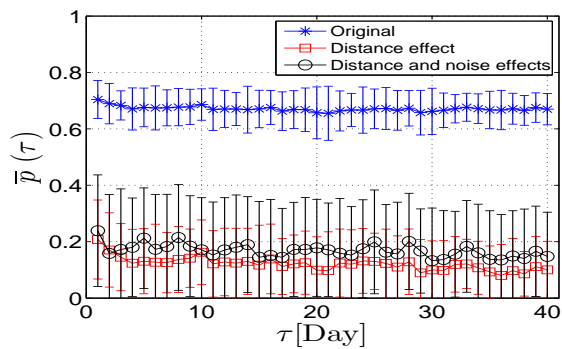}\label{tau_all_fillter_b}}
        \end{center}
        \caption{The average correlation, $\overline{p}(\tau)$, between network
       	adjacency matrices at different time snapshot, between $\tau=1$ and
       	$\tau=40\ years$ apart, for networks based on temperature at 850hPa and
       	located at (a) zone 1, and (b) zone 9. The upper curve in each figure,
        represents the correlation between the original networks without
        removing the effects of distance and noise. The lower curve
        represents $\overline{p}(\tau)$ for a network after removing of the
       	distance effect. The mid curve represents $\overline{p}(\tau)$ for a
       	network after removing both, distance and noise
       	effects.\label{tau_all_fillter}}
    \end{figure}

	In Sec.~\ref{sec:single} we showed that single links remain relatively stable.
	In this section we study the stability of the entire hierarchy of links within
	the climate network. We find the network to be relatively stable over time.
	This stability of the network may be demonstrated by measuring the similarity between network
	states in different years. We analyze the similarity by calculating ${p}(\tau)$
	(the Pearson coefficient) between the adjacency matrices of two network states
	in different years $y_1,y_2$ as a function of $\tau = y_2-y_1$. In
	Fig.~\ref{tau_all_fillter} (the upper curve) we show the average similarity
	$\overline{p}(\tau)$ between network structures as a function of the time
	separation $\tau$. It is seen in Fig.~\ref{tau_all_fillter_a} that this
	similarity is indeed high and almost constant, $\overline{p}(\tau)\approx0.8$.
	This behavior is consistent for all networks in the non-equatorial regions. For
	networks in equatorial regions the correlation between the network states in
	different years, $\overline{p}(\tau)$ is still significantly high but relatively
	fluctuative. (see Fig.~\ref{tau_all_fillter_b}).

	In Fig.~\ref{distance_profile_eq} we showed, that there is a strong dependence
	between the link strength, $W$ and the link distance, $D$. Links with
	shorter distances $D$ are therefore more likely to have higher $W$ values, at all
	times. It is therefore plausible that the high stability of
	$\overline{p}(\tau)\approx0.8$ is partially due to this strong dependence. The
	contribution of the effect of the $W$--$D$ dependence to this observed
	stability, on the one hand, and the contribution of physical coupling
	processes, on the other hand, must be estimated.
	
	We achieve this goal of removing the contribution of the $W$--$D$ dependence
	by subtracting from each link strength, $W_{l,r}$ the average strength of the
	group of links with a similar distance, $\overline{W}_D$. A new, transformed
	adjacency matrix, $W_{l,r}-\overline{W}_D$, is thus formed, which does not
	depend on $D$. Repeating our analysis of calculating $\overline{p}(\tau)$, for
	the new adjacency matrix, we show in the lower curve of
	Fig.~\ref{tau_all_fillter}, the stability of this network. Indeed, after
	removal of the distance effect, the network exhibits lower $\overline{p}(\tau)$
	values. However, the stability related to physical coupling processes is still
	significant~\footnote{Similar analysis with shuffled data yields values which
	are smaller by typically a factor of \protect{$10$}}.
	
	It is plausible that some of our network links emerge mainly due to noise and
	not due to real physical coupling processes. In Sec.~\ref{sec:false_true} we
	show that this group of false links is characterized by low $W_{l,r}$ values at
	all times, and that this characterization is sufficient for uniquely
	identifying this group. In order to identify this group of false links which
	are due to noise, we define $\overline{W}_{l,r}$ as the average strength of a
	link $W^y_{l,r}$ over all years. Upon eliminating low weighted links that satisfy
	$\overline{W}_{l,r} < \theta$ (where $\theta$ is a threshold that will be
	determined later in Sec.~\ref{sec:false_true}) from our network, we observe in
	the middle curve of Fig.~\ref{tau_all_fillter} an increase of the network
	stability. Thus, while the hierarchy of false links rapidly changes in each
	time step, the hierarchy of significant links (having
	$\overline{W}_{l,r}>\theta$) is, to a large extent, preserved.
	
	The new $\overline{p}(\tau)$ values, which are calculated after the removal of
	both the distance effect and the effects of noise, for different regions and
	fields are summarized in Table ~\ref{corr_table}. In contrast to the common
	$\overline{p}(\tau)=0.8$ value that was observed for the original network
	(including the distance and noise effects), after removal of the distance and
	noise effects we observe lower $\overline{p}(\tau)$ values specific to each zone and
	climate variable. Still, in general, non equatorial regions exhibit larger
	stability values than equatorial regions.
	
	It has been shown that during El-Ni\~{n}o times, link strengths, $W^y_{l,r}$, 
	are significantly reduced mainly in equatorial regions~\cite{survival,EBauton}.
	Hence our observation of lower stability in equatorial regions is
	consistent with the known effect of El-Ni\~{n}o on the climate network.
		
	From Table~\ref{corr_table} we see that removing both the distance effect and
	the effects of noise reveal that the networks in zone $3$, at the southern ocean,
	exhibit low stability, $\overline{p}(\tau)$ values similar to the equatorial regions.
	This similarity of the behavior of the network in zone $3$ and the behavior of
	the network in equatorial regions is consistent with the known
	local oscillations in zone $3$ that correlate with ENSO, due to both ocean
	mechanisms \cite{whi96} and atmospheric coupling mechanism
	~\cite{Jiping}.

	Based on the high stability values seen in Table ~\ref{corr_table}, we conclude
	that similarity between network states at all times stems from a hierarchy of
	real physical correlations (links) between different locations, which is
	preserved in time.

	\begin{table}
     	\begin{tabular}{|c||c|c|c|c|}
 		\hline
 		\multirow{2}{*} \protect{\backslashbox{\\climate\\ variable}{zone}} &  Temp &
 		Temp & Height  & Height \\ &   850 hPa &  500 hPa &  850 hPa &  500 hPa\\ \hline
 		\hline
 		zone 1 & $0.42\pm0.08$ & $0.31\pm0.08$ & $0.48\pm0.08$ & $0.3\pm0.1$\\
 		\hline
 		zone 2 & $0.4\pm0.08 $ & $0.33\pm0.08 $ & $0.32\pm0.11 $ & $0.3\pm0.11$\\
 		\hline        
 		zone 3 & $0.27\pm0.09 $ & $0.21\pm0.1 $ & $0.17\pm0.13 $ & $0.18\pm0.11$\\
 		\hline
 	 	zone 4 & $0.32\pm0.08 $ & $0.24\pm0.09 $ & $0.45\pm0.11 $ & $0.36\pm0.1$\\
 	 	\hline
 	 	zone 5 & $0.42\pm0.1 $ & $0.34\pm0.1 $ & $0.51\pm0.12 $ & $0.42\pm0.11$\\
 	 	\hline
 	 	zone 6 & $0.43\pm0.08$ & $0.41\pm0.09 $ & $0.45\pm0.08 $ & $0.37\pm0.08$\\
 	 	\hline
 	 	zone 7 & $0.22\pm0.11 $ & $0.37\pm0.15 $ & $0.37\pm0.16 $ & $0.18\pm0.17 $\\
 	 	\hline
 	 	zone 8 & $0.33\pm0.13 $ & $0.19\pm0.14 $ & $0.24\pm0.15 $ & $0.14\pm0.21 $\\
 	 	\hline
 	 	zone 9 & $0.25\pm0.14 $ & $0.2\pm0.11 $ & $0.37\pm0.14 $ & $0.16\pm0.2 $
 	    \end{tabular}
 		\caption{The average correlation values, $\overline{p}(\tau)$, between
 		network adjacency matrices at different time snapshot for networks based on
 		various fields and located at different regions. The values shown are
 		after removing the distance and noise effects.
 		\label{corr_table}}
    \end{table}

\section{\label{sec:Similarity}Similarity between the networks structure, in
different altitudes and different climate variables}
	
	A further indication that the stability of the network structure reflects
	a stability of physical coupling processes, is from the finding of 
	similarity between the networks structure in different altitudes, and different
	climate variables. For example, synchronized heating of two sites at the 850hPa
	isobar network is likely to also cause synchronized heating of the
	corresponding sites in the adjacent isobar of 500hPa network by direct heat
	transport.
	In the following we show that such a correspondence
	between networks of different altitudes and climate variables indeed exists.
	
\begin{figure}[ht]
    	\begin{center}
        	\subfigure[\ Height-850-500hPa, zone1]{\includegraphics[width =
        	8cm,height =5cm]{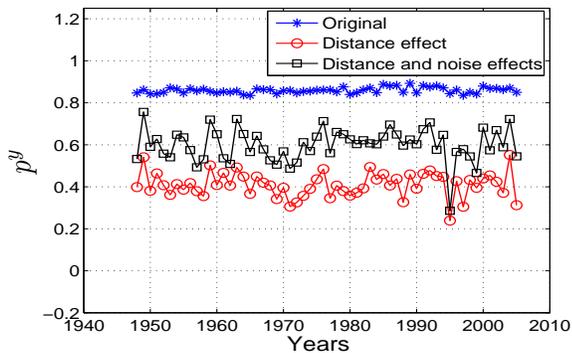}\label{crossNetworks_a}}
	        \subfigure[\ Height-850-500hPa, zone9]{\includegraphics[width =
            8cm,height
            =5cm]{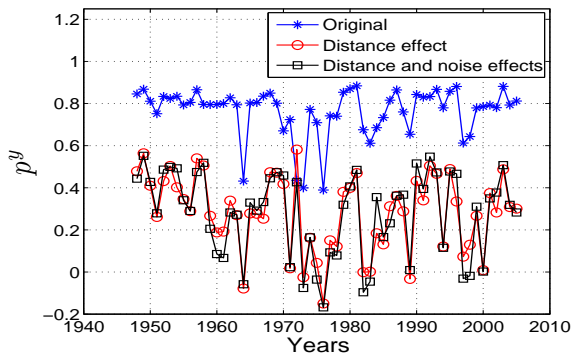}\label{crossNetworks_b}}
        	\end{center}
        	\caption{The correlation between two network adjacency matrices at
 		different altitudes, 850hPa isobar and 500hPa isobar in equatorial and
 		non--equatorial regions. (a) Zone 1 (non--equatorial region), and (b) zone 9
 		(equatorial region). The upper curve in each figure, represents the
 		correlation between the original networks without removing effects of
 		distance and noise. The lower curve represents ${p^y}$ for the networks
 		after removing the distance effect. The mid curve represents
 		${p^y}$ for the networks after removing both, distance and noise
 		effects.\label{crossNetworks}}
    \end{figure}

	In Fig.~\ref{crossNetworks} we show the Pearson correlation $p^y$ between the
	adjacency matrices of the climate networks in the 850hPa isobar and in the
	500hPa as a function of time. Similar to Sec.~\ref{sec:Stability} the upper
	curve represents the values of $p^y$ for the original networks, the lower curve
	is after the removal of the distance effect, and the middle curve is after the
	removal of both distance and noise effects. As is clearly seen from
	Fig.~\ref{crossNetworks_a} this correlation is significant, with an average
	value of $\overline{p^y}\approx0.6$ and small fluctuation during time. In all
	non--equatorial (for both temperature and geopotential height) regions we
	observe similar high $\overline{p^y}$ values. A further observation from
	Fig.~\ref{crossNetworks_a} is the increase of $10$--$20\%$ in $\overline{p^y}$
	values after the removal of noise.
	
	In contrast, in equatorial regions (see e.g. Fig.~\ref{crossNetworks_b}) we
	generally observe lower and more fluctuative ${p^y}$ values.
	Another difference between non--equatorial and equatorial regions is the
	smaller effect of the removal of noise on the  ${p^y}$.
	
\begin{figure}[ht]
    	\begin{center}
    		\includegraphics[width = 8cm,height =
    	5cm]{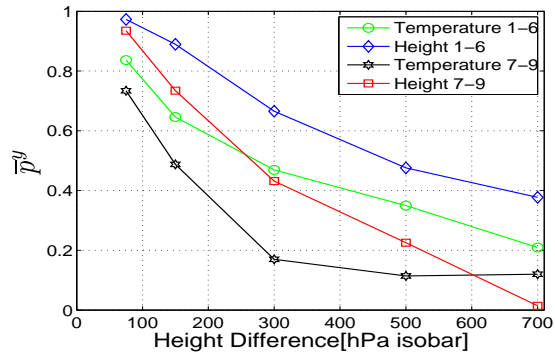}
       	\end{center}
       	\caption{Average correlations between pairs of networks with different
       	altitude distances. The four curves describe two networks which are
       	located in non--equatorial regions (zones 1--6) and based on geopotential height
       	or based on temperature and two networks which are located in
       	equatorial regions (zones 7--9) and based on geopotential height or
       	based on temperature measurements.
       	\label{crossNetworksDiffAltitude}}
    \end{figure}

    Based on physical considerations it is reasonable that a pair of networks with
    a larger altitude distance will have a smaller similarity. Indeed such a
    monotonic decreasing relation is seen in
   	Fig.~\ref{crossNetworksDiffAltitude}. Each point in the curves of
   	Fig.~\ref{crossNetworksDiffAltitude} is an average over different regions,
   	different time snapshots, and different altitudes of the correspondence
   	$\overline{p^y}$. The different curves indicates that this monotonic
   	decrease in the similarity behavior holds both in equatorial and
   	non--equatorial regions and both for temperature and geopotential height networks.
	
\subsection{\label{sec:false_true}Criterion for significant links}

	Underlying our supporting arguments for the stability of the climate network,
	there is an assumption. We rely on the existence of a sharp boundary between the
	properties of links that result due to noise, $L_{N}$, and links
	that result due to real physical dependence, $L_{p}$. The set of all links
	is $L=L_{N}\cup\ L_{p}$. In this section we will show that such a boundary
	indeed exists, with respect to two link properties: \textbf{(a)} the average
	over all years of the link strength , $\overline{W}_{l,r}$ and \textbf{(b)} the
	variability over time of the time delays of links $STD(T_{l,r})$. We will
	later show that using both quantities in order to identify the set of real
	links $L_{p}$ converges to almost the same set of links.

	Our anchor for comparison between the derived $L_{p}$ and $L_{N}$ is the
	distributions of \textbf{(a)} $\overline{W}_{l,r}$ and \textbf{(b)}
	$STD(T_{l,r})$ for networks based on shuffled data. The shuffling scheme is
	aimed at preserving all the statistical quantities of the data, such as the
	distribution of values, and their autocorrelation properties, but omitting the
	physical dependence between different nodes (different geographical locations).
	The network properties in such a case are only due to the statistical
	quantities and therefore are similar in their properties to false links. To
	achieve this shuffling goal, we choose for each node a random sequence of $y$
	in $S^y_d$ (the order of the days, $d$ is preserved) ~\footnote{When cross correlations
	are calculated, the continuation of the same year \protect{$y$} is used, rather
	than the next random sequence}. Thereafter, the entire construction of
	the network, based on correlations of the shuffled records, is performed. The
	adjacency matrix of the network based on shuffled data is denoted as $w_{l,r}$.
	The time delay matrix of the network based on shuffled data is denoted as
	$t_{l,r}$.

\begin{figure}[ht]
		\subfigure[\ Temp-850hPa, zone1]{\includegraphics[width = 0.5\textwidth,height
		= 5cm]{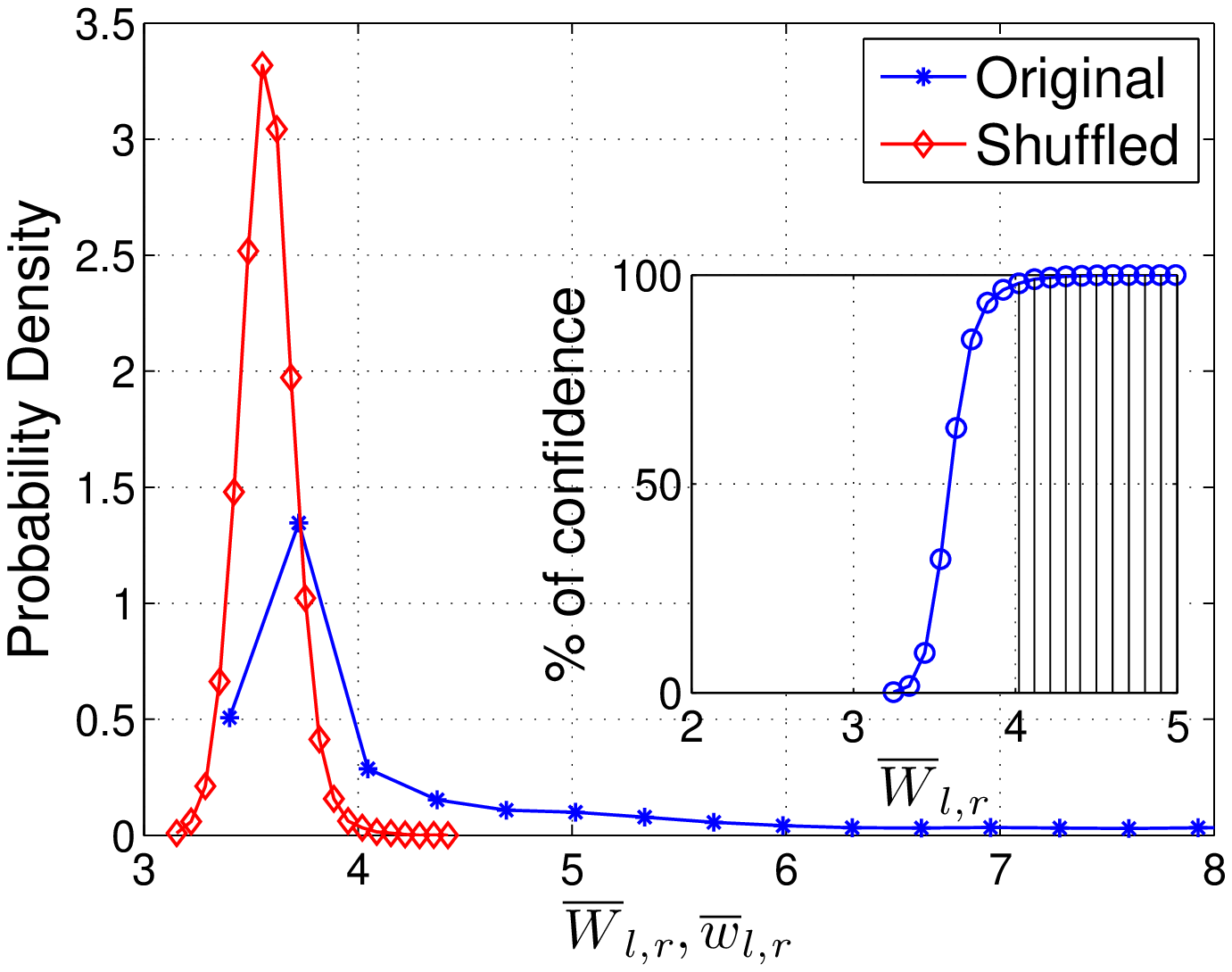}\label{str_confidence_a}}
	 	\subfigure[\ Temp-850hPa, zone9]{\includegraphics[width =
	 	0.5\textwidth,height =5cm]{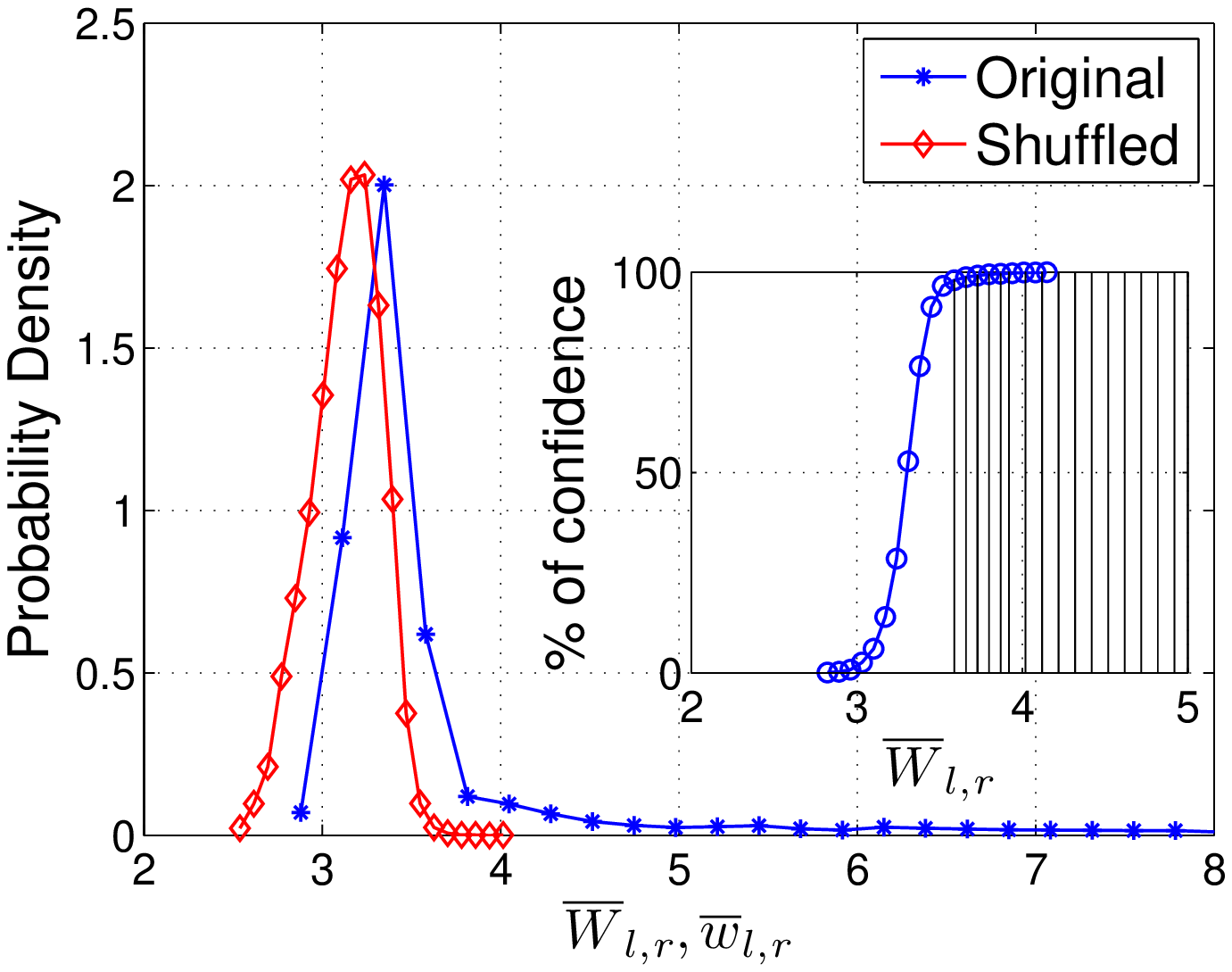}\label{str_confidence_b}}
      	\caption{The distribution of $\overline{{W}}_{l,r}$ and
      	$\overline{{w}}_{l,r}$ in equatorial and
 		non--equatorial regions for networks based on temperature measurements at
 		850hPa isobar. (a) Zone 1 (non--equatorial region), and (b) zone 9
 		(equatorial region).
      	\label{str_confidence}}
    \end{figure}

	\textbf{(a) Average link strength.} In Fig.~\ref{str_confidence} we compare the
	probability density function (PDF) of $\overline{W}_{l,r}$ and
	$\overline{w}_{l,r}$. As clearly seen from these figures, the range of possible
	$\overline{w}_{l,r}$ is extended only over a limited range of values,
	$\overline{w}_{l,r}\in [3,4]$. Higher values that exist in the PDF for
	$\overline{W}_{l,r}$ are missing from the PDF of the shuffled data, and
	therefore are not likely to occur by chance. The cumulative distribution
	function (CDF) of $\overline{w}_{l,r}$ (see insets of
	Fig.~\ref{str_confidence}) can be regarded as an estimate for the likelihood of a $\overline{W}_{l,r}$
	value to arise by real physical dependence. The $98\%$ likelihood level is
	shaded in the inset of Fig.~\ref{str_confidence_a}, having $\overline{W}_{l,r}
	\geq 4$ for off-equatorial regions, and in the inset of
	Fig.~\ref{str_confidence_b}, $\overline{W}_{l,r} \geq 3.6$ for equatorial
	regions.
	
\begin{figure}[ht]
		\subfigure[\ Temp-850hPa, zone1]{\includegraphics[width = 8cm,height =
	 	5cm]{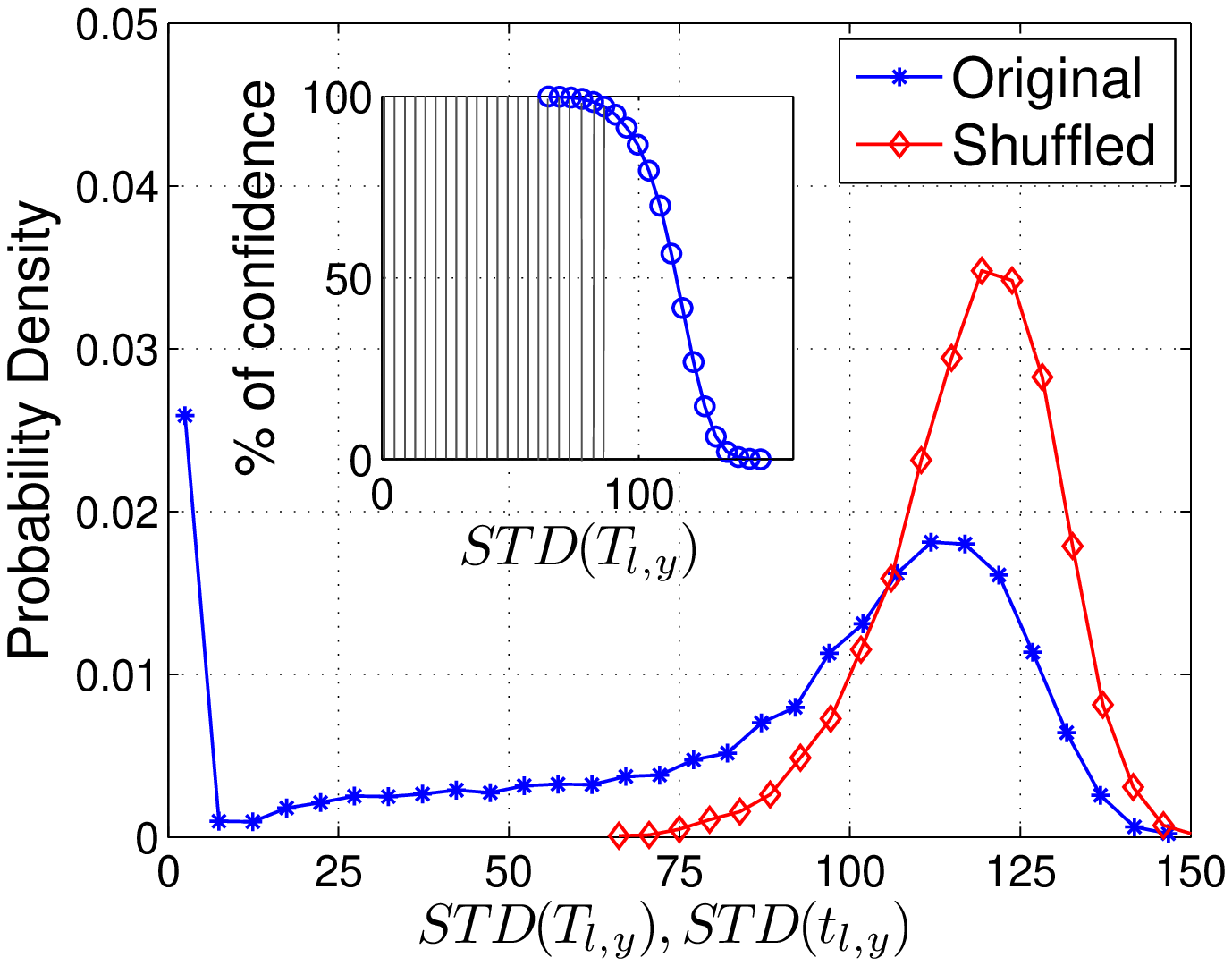}}
	 	\subfigure[\ Temp-850hPa, zone9]{\includegraphics[width = 8cm,height = 5cm]
	 	{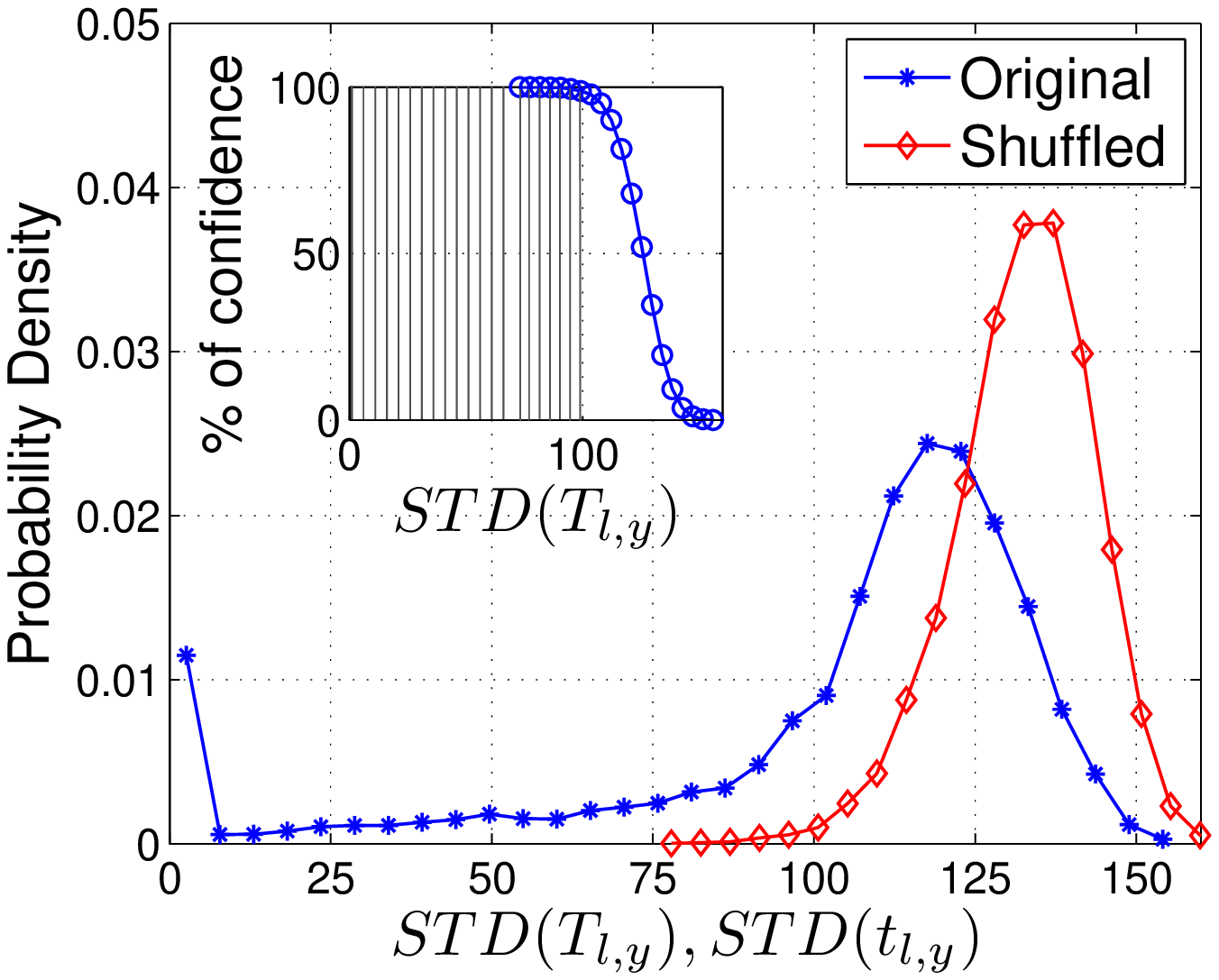}}
		\caption{The distribution of $STD({{T}_{l,r}})$ and
      	$STD({{t}_{l,r}})$ in equatorial and
 		non--equatorial regions for networks base on temperature measurements at
 		850hPa isobar. (a) Zone 1 (non--equatorial region), and (b) zone 9
 		(equatorial region).
      	\label{time_delay_confidence}}
    \end{figure}

	\textbf{(b) Variation of the time delay.} High variability during different
	years of the time delays of links $STD(T_{l,r})$, is also a signature of
	artificial (random) behavior~\footnote{The stability of the time delay with
	time as a criterion for a real link was proposed for physiological networks by
	A.Bashan R. Bartsch, J. Kantelhardt, S. Havlin, and P. Ivanov, Physiological
	Networks: towards systems physiology, Preprint (2011)}. Therefore it serves as
	another good separator between $L_{p}$ and $L_{N}$. In
	Fig.~\ref{time_delay_confidence} we show the probability density function (PDF)
	of $STD(T_{l,r})$ and $STD(t_{l,r})$. As clearly seen from these figures, the
	range of possible $STD(t_{l,r})$ is extended over a limited range of values,
	$STD(t_{l,r})\in [75,150]$. Lower values that exist in the PDF of
	$STD(T_{l,r})$ are missing from the PDF of the shuffled data, and therefore are
	not likely to arise by chance. The cumulative distribution function (CDF) of
	$STD(t_{l,r})$ (see inset of Fig.~\ref{time_delay_confidence}) can be regarded
	as an estimate for the likelihood of a $STD(T_{l,r})$ value to occur by real
	physical dependence. The $98\%$ likelihood level is shaded in the inset of
	Fig.~\ref{time_delay_confidence}, having $STD(T_{l,r}) \leq 75$ for
	off-equatorial regions, and $STD(T_{l,r}) \leq 80$ for equatorial regions.

\begin{figure}[ht]
		\includegraphics[width =
		0.45\textwidth,angle=270]
		{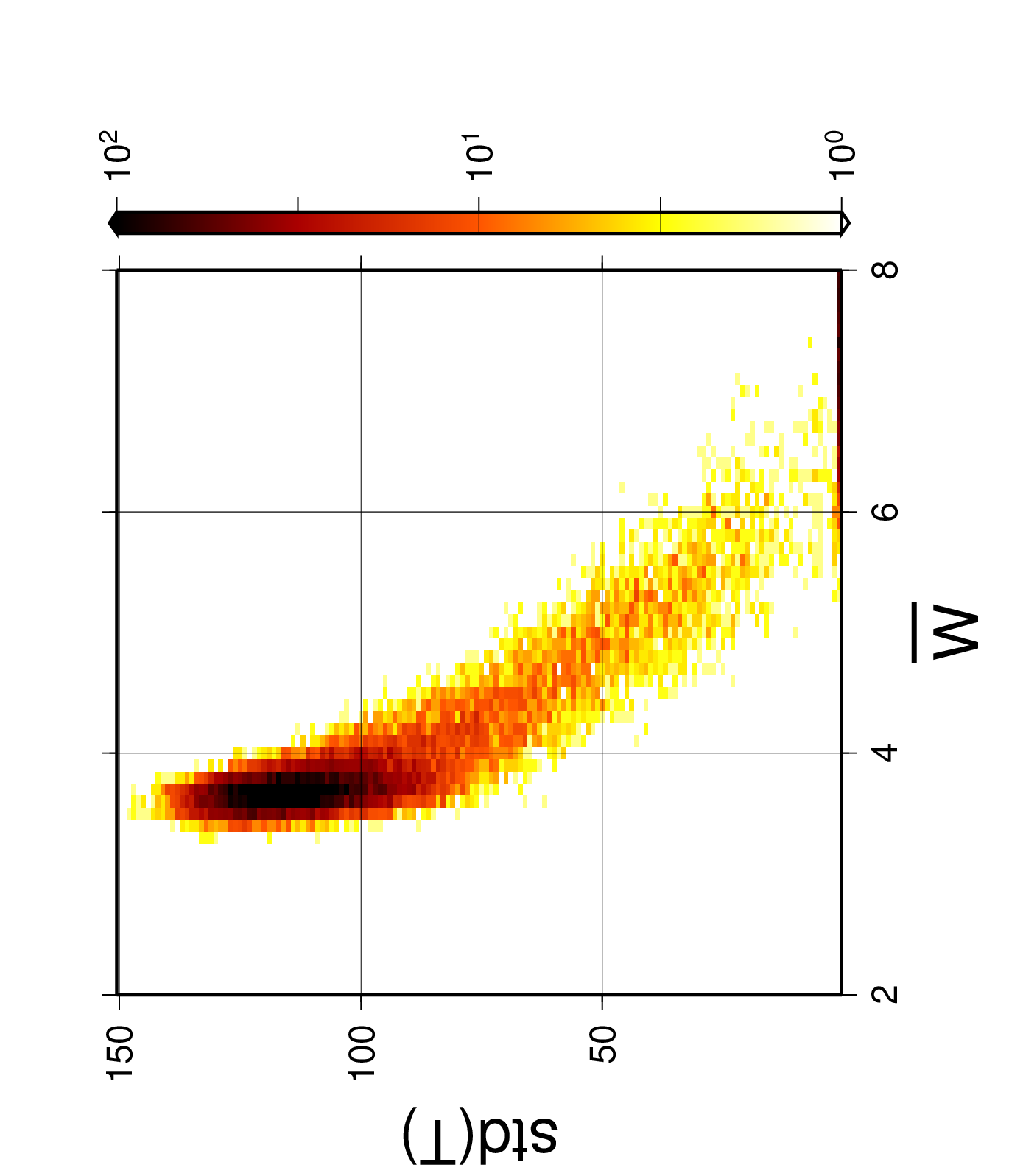} \caption{(color online). The 2D histogram of links time delay
		variation, $STD(T_{l,r})$ and links average strength, $\overline{W}_{l,r}$.
		This 2D histogram is for network located at zone 1 and based on temperature
		measurements at 850hPa isobar.
    	\label{time_delay_vs_str}}
	\end{figure}

	Both $\overline{W}_{l,r}$ and $STD(T_{l,r})$ can be used for determining a
	boundary between $L_P$ and $L_N$. Convergence to similar $L_P$ and $L_N$ in
	both criterions can be considered as a confirmation that either of these
	criterions indeed efficiently distinct between links that emerge merely due
	to noise and real links. In Fig.~\ref{time_delay_vs_str} we show a two
	dimensional PDF of $\overline{W}_{l,r}$ and $STD(T_{l,r})$. A large fraction of
	the links evidently have both low values of $\overline{W}_{l,r}$ and high
	values of $STD(T_{l,r})$, which is a typical behavior of links that emerge from
	random behavior. This set of links is realized as a sharp local maximum of the
	PDF in the region $\overline{W}_{l,r}\in\ \left[3,4.5\right], STD(T_{l,r})\in\
	\left[75,150\right]$ (See Figs.~\ref{str_confidence} and
	~\ref{time_delay_confidence}). Within this region, $\overline{W}_{l,r}$ and
	$STD(T_{l,r})$ are not correlated, since the fluctuations are random in both
	axes. Outside this region the mutual local maximum of the PDF in both axis are
	correlated, i.e. larger values of $\overline{W}_{l,r}$ are paired with lower
	values of $STD(T_{l,r})$. The crossover between the two regimes occurs around
	$\left(\overline{W}_{l,r}, STD\left(T_{l,r}\right)\right)=(4.5,75)$. This
	qualitative behavior is consistent at all regions (1--9), but the crossover
	point is a bit different for non--equatorial regions (around
	$\left(4.0,75\right)$).

\begin{figure}[ht]
 		\includegraphics[width =
 		0.4\textwidth]{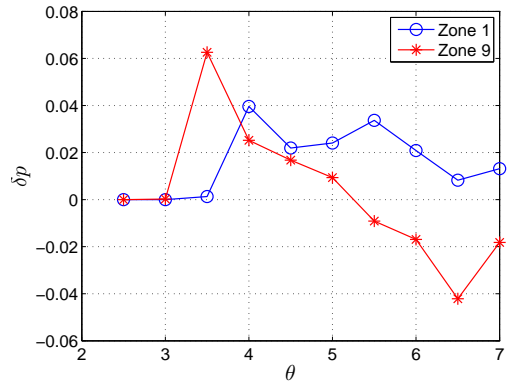} 
 		\caption{(color online). The sensitivity, $\delta p$, of the stability
		function $p$ as a function of the threshold $\theta$. 
     	\label{optimization}}
 	\end{figure}

 	A further indication that the boundary between $L_{p}$ and $L_{N}$ is within
 	the region $\overline{W}_{l,r}\in\ \left[3.5,4.5\right]$ is the increased
 	sensitivity of the stability measure $\overline{p}\left(\tau;\theta\right)$ to
 	the removal of noise within this region. Here we indicate explicitly the value
 	of the threshold for noise removal by the second argument $\theta$. In
 	Fig.~\ref{optimization} we show the differential of $\overline{p}\left(\tau;\theta\right)$, averaged over all values of $\tau$,
 	$\delta
 	p\equiv\overline{p}\left(\theta+\delta\theta\right)-\overline{p}\left(\theta\right)$,
 	where $\delta \theta=0.5$. We find a sharp local maximum in
 	$\delta{p}$, around $\overline{W}_{l,r}\in\ \left[3.5,4.5\right]$.
 	This sharp maximum is consistent with $\theta$ crossing the boundary
 	between $L_{N}$ and $L_{p}$,	 where many of the links related to noise drop
 	off the network, and causes the average stability
 	$\overline{p}\left(\theta\right)$ to abruptly rise. Such behavior of $\delta
 	p$ is consistent both in equatorial regions (stars) and non--equatorial
 	regions (circles). The response of the sensitivity $\delta p$ to further
 	removal of links (which mainly belong to $L_{p}$) is thereafter reduced. In
 	fact, removal of physical links might even result in a reduction of the
 	stability (e.g. $\delta p<0$), as is indeed observed for large $\theta$ in the
 	equatorial curve in Fig.~\ref{optimization}. In conclusion, a sharp boundary
 	between $L_{N}$ and $L_{p}$ is almost certainly identified around
 	$\overline{W}_{l,r}\approx 4.0\pm 0.5$ in the networks calculated for all
 	types of data (temperature and geopotential height in various altitudes
 	covering the troposphere), both in equatorial and non--equatorial regions.
 	 
\section{\label{sec:Summary}Summary}
	In summary, we have established the stability of the network of connections
	between the dynamics of climate variables (e.g. temperatures and geopotential
	heights) in different geographical regions. The strength of the physical
	connection, $W_{l,r}$, that each link in this network represents, changes only
	between $5\%$ to $30\%$ over time. A clear boundary between links that
	represent real physical dependence and links that emerge due to noise is shown to exist.
	The distinction is based on the high link strength $W_{l,r}$ and on the
	variability of time delays $STD\left(T_{l,r}\right)$.

	Beside the stability of single links, also the hierarchy between the link
	strengths is preserved to a large extent. We have shown that this hierarchy is
	partially due to the two dimensional space in which the network is embedded,
	and partially due to physical coupling processes. Moreover the contribution of
	each of these effects, and the level of noise was explicitly estimated. The
	spatial effect is typically around $50\%$ of the observed stability, and the
	noise reduces the stability value by typically $5\%$--$10\%$.
	
	The network structure was further shown to be consistent across different
	altitudes, and a monotonic relation between the altitude distance and the
	correspondence between the network structures is shown to exist. This yields
	another indication that the observed network structure represents effects of
	physical coupling.

	The stability of the network and the contributions of different effects were
	summarized in specific relation to different geographical areas, and a clear
	distinction between equatorial and non--equatorial areas was observed.
	Generally, the network structure of equatorial regions is less stable and more
	fluctuative.
	
	The stability and consistence of the network structure during time and across
	different altitudes stands in contrast to the known unstable variability of the
	daily anomalies of climate variables. This contrast indicates an analogy
	between the behavior of nodes in the climate network and the behavior of
	coupled chaotic oscillators ~\cite{SCO_review}.
	
	A future outreach of our work can be a mapping between network features (such
	as network motifs) and known physical processes. Such a mapping was previously
	shown to exist~\cite{EBauton} between an autonomous cluster in the climate
	network and El-Ni\~{n}o. Further structures without such a climate
	interpretation might point towards physical coupling processes which were not
	observed earlier.
 

\end{document}